\documentclass{appolb}

\usepackage{amsmath}
\usepackage{amsfonts}
\usepackage{bbm}

\begin{document}
\newcommand{\NN}{\textrm{N}}
\newcommand{\rmd}{\mathrm{d}}

\title{Wave functions of linear systems}
\author{Tomasz Sowi\'nski
\address{Center for Theoretical Physics \\ of the Polish Academy of Sciences\\ Al. Lotnik\'ow 32/46, Warsaw, Poland\\tomsow@cft.edu.pl}
}
\maketitle
\begin{abstract}
Complete analysis of quantum wave functions of linear systems in an arbitrary number of dimensions is given. It is shown how one can construct a complete set of stationary quantum states of an arbitrary linear system from purely classical arguments. This construction is possible because for linear systems classical dynamics carries the whole information about quantum dynamics.  
\end{abstract}
  
\section{Introduction}
Connection between classical and quantum description of physical system manifests itself in very amazing and nontrivial way. We understand how to describe any system in classical and quantum language and also we believe that quantum description should smoothly transform to classical one when we neglect quantum corrections\footnote{It is often said that we go with $\hbar$ to zero.}. But in general we still do not understand how one can make this transformation in practice and identity the classical motion in quantum one. The first description how one should understand this correspondence was given by Ehrenfest in his famous theorem \cite{Ehrenfest}. It states that the expectation values of physical observables evolve in time almost as classical quantities (up to quantum corrections). In the case of linear systems they evolve exactly in the same way. This is evident in the Heisenberg picture of quantum evolution. Since in the case of a linear system classical Hamiltonian equations of motion are linear they have exactly the same structure as quantum Heisenberg equations of evolution of quantum operators. Therefore, in this case the classical dynamics carries the full information about quantum dynamics. 

The question which we answer in this paper is how this exact connection between classical and quantum dynamics in arbitrary number of dimmensions is realized in the Schr\"odinger picture where the whole dynamics is contained in the evolution of the wave function of the system. Full discussion of this correspondence in one dimensional case was given before in \cite{Andrews}.

The main observation presented here is a multidimensional generalization of our idea proposed earlier in \cite{TSIBB} for the three dimensional case.

\section{Class of linear systems}
For each $\NN$-dimensional linear system there always exist such canonical variables $\boldsymbol{\xi}$ and $\boldsymbol{\pi}$ that the system is described by the Hamiltonian of the following form
\begin{equation} \label{hamo1}
{\cal H} = \frac{1}{2m}\,\boldsymbol{\pi}\!\cdot\!\hat{F}\!\cdot\!\boldsymbol{\pi} +
\boldsymbol{\xi}\!\cdot\!\hat{Q}\!\cdot\!\boldsymbol{\pi}+
\frac{m}{2}\boldsymbol{\xi}\!\cdot\!\hat{U}\!\cdot\!\boldsymbol{\xi}+
m\,\boldsymbol{f}(t)\!\cdot\!\boldsymbol{\xi}+\frac{1}{m}\boldsymbol{h}(t)\!\cdot\!\boldsymbol{\pi},
\end{equation}
where the matrices $\hat{F}$ and $\hat{U}$ are symmetric and the vectors $\boldsymbol{f}(t)$ and $\boldsymbol{h}(t)$ are given functions of time. If $\boldsymbol{\pi}$ is chosen as a kinetic momentum then the matrix $\hat{F}$ should be positive definite (kinetic energy should increase with momentum of the particle). Therefore there exists a matrix $\hat{O}$ which diagonalizes the matrix $\hat{F}$ to identity
\begin{equation}
\hat{O}^{\mathrm{T}}\!\cdot\!\hat{F}\!\cdot\!\hat{O} = \hat{I}.
\end{equation}
Let us define the following matrices
\begin{subequations}
\begin{eqnarray}
\hat{W} &=& \hat{O}^{-1}\!\cdot\!\hat{Q}\!\cdot\!\hat{O}, \\
\hat{S} &=& \frac{1}{2}\left(\hat{W}+\hat{W}^{\mathrm{T}}\right), \\
\hat{\Omega} &=& \frac{1}{2}\left(\hat{W}-\hat{W}^{\mathrm{T}}\right),
\end{eqnarray}
\end{subequations}
and now let us make the following canonical transformation to the new canonical variables $\boldsymbol{r}$ and $\boldsymbol{p}$
\begin{subequations}
\begin{eqnarray}
\boldsymbol{r} &=& \hat{O}^{\mathrm{T}}\!\cdot\!\boldsymbol{\xi}, \\
\boldsymbol{p} &=& \hat{O}^{-1}\!\cdot\!\boldsymbol{\pi}+m\hat{S}\!\cdot\!\hat{O}^{\mathrm{T}}\!\cdot\!\boldsymbol{\xi}+\left(\hat{O}^{\mathrm{T}}\right)^{-1}\!\cdot\!\boldsymbol{h}(t).
\end{eqnarray}
\end{subequations}
One can show that it is indeed a canonical transformation because the Poisson bracket structure is preserved. This transformation is in fact the composition of three simpler canonical transformations: the diagonalization of the kinetic energy term
\begin{subequations}
\begin{equation}
\boldsymbol{r}'=\hat{O}^{\mathrm{T}}\cdot\boldsymbol{\xi},\qquad \boldsymbol{p}'=\hat{O}^{-1}\cdot\boldsymbol{\pi},
\end{equation}
the shift in the momentum space
\begin{equation}
\boldsymbol{r}''=\boldsymbol{r}',\qquad \boldsymbol{p}''=\boldsymbol{p}'+\left(\hat{O}^{\mathrm{T}}\right)^{-1}\cdot\boldsymbol{h}(t),
\end{equation}
and the stretching transformation of the phase space of the system
\begin{equation}
\boldsymbol{r}=\boldsymbol{r}'',\qquad \boldsymbol{p}=\boldsymbol{p}''+m\hat{S}\cdot\boldsymbol{r}''.
\end{equation}
\end{subequations}
In our new variables, the Hamiltonian of the system has the form
\begin{equation} \label{hamiltonian}
{\cal H}=\frac{\boldsymbol{p}^2}{2m} + \boldsymbol{r}\!\cdot\!\hat{\Omega}\!\cdot\!\boldsymbol{p} + \frac{m}{2} \boldsymbol{r}\!\cdot\!\hat{V}\!\cdot\!\boldsymbol{r} - m\boldsymbol{g}(t)\!\cdot\!\boldsymbol{r}-\frac{1}{2m}\left[\left(\hat{O}^{\mathrm{T}}\right)^{-1}\!\cdot\!\boldsymbol{h}(t)\right]^{2},
\end{equation}
where
\begin{subequations}
\begin{eqnarray}
\hat{V} &=& \hat{O}^{-1}\!\cdot\!\hat{U}\!\cdot\!\left(\hat{O}^{T}\right)^{-1}-\hat{S}^{2}-\left[\hat{\Omega},\hat{S}\right], \\
\boldsymbol{g}(t) &=& \frac{1}{m}\hat{W}\!\cdot\!\left(\hat{O}^{\mathrm{T}}\right)^{-1}\!\cdot\!\boldsymbol{h}(t)-\hat{O}\!\cdot\!\boldsymbol{f}(t).
\end{eqnarray}
\end{subequations}
The last term in the Hamiltonian (\ref{hamiltonian}) depends only on time. Therefore, without loss of generality we can omit it. 
\section{Classical dynamics}
Classical equations of motion following from the Hamiltonian (\ref{hamiltonian}) have the form
\begin{subequations} \label{eqmotion}
\begin{eqnarray}
\frac{\rmd \boldsymbol{r}}{\rmd t} &=& \frac{\boldsymbol{p}}{m}-\hat{\Omega}\!\cdot\!\boldsymbol{r}, \\
\frac{\rmd \boldsymbol{p}}{\rmd t} &=& -m\hat{V}\!\cdot\!\boldsymbol{r}-\hat{\Omega}\!\cdot\!\boldsymbol{p} - m\boldsymbol{g}(t).
\end{eqnarray}
\end{subequations}
Of course the best way to solve these equations is to solve first a homogeneous 
equation with $\boldsymbol{g}(t)\equiv 0$. In that case any solution of equations (\ref{eqmotion}) can be expressed as a sum of the eigenmodes
\begin{equation}
\left( \begin{matrix}
\boldsymbol{r}(t) \\[3pt] \boldsymbol{p}(t) \end{matrix} \right) = \sum_{k=1}^{2\NN} \lambda_k \left( \begin{matrix} \boldsymbol{R}_k \\[3pt] \boldsymbol{P}_k \end{matrix} \right) \mathrm{e}^{i\omega_k t},
\end{equation}
where the coefficients $\lambda_k$ are determined by initial conditions. The characteristic frequencies $\omega_k$ and the amplitudes $\left(\boldsymbol{R}_k,\boldsymbol{P}_k\right)$ obey the following matrix equation
\begin{equation} \label{charpoly}
\left( \begin{matrix} -\hat{\Omega}-i\omega_k & \frac{1}{m} \\[3pt] -m\hat{V} & -\hat{\Omega}-i\omega_k
\end{matrix} \right) \cdot \left(\begin{matrix} \boldsymbol{R}_k \\[3pt] \boldsymbol{P}_k \end{matrix} \right) = 0
\end{equation}
In the Appendix \ref{appmatrix} we show that for such a matrix, the eigenvalues appear in pairs. It means that for an $\NN$ dimensional system we always have the following frequencies
\begin{equation} 
\pm \omega_1,\; \pm \omega_2, \; \ldots , \;\pm \omega_{\NN-1},\; \pm \omega_\NN.
\end{equation}
This means that amplitudes of the modes $\omega_i$ and $-\omega_i$ are related by complex conjugation. Therefore, any physical solution of Eqs. (\ref{eqmotion}) can be expressed in the following way
\begin{equation} \label{physsolution}
\left( \begin{matrix}
\boldsymbol{r}(t) \\[3pt] \boldsymbol{p}(t) \end{matrix} \right) = \sum_{i=1}^{\NN} \left[\lambda_i \left( \begin{matrix} \boldsymbol{R}_i \\[3pt] \boldsymbol{P}_i \end{matrix} \right) \mathrm{e}^{i\omega_i t} + \lambda_i^* \left( \begin{matrix} \boldsymbol{R}_i^* \\[3pt] \boldsymbol{P}_i^* \end{matrix} \right) \mathrm{e}^{-i\omega_i t}\right].
\end{equation}
In the presence of the external field the only difference is that the parameters $\lambda_i$ are time dependent and the general solution has a form
\begin{equation} 
\left( \begin{matrix}
\boldsymbol{r}(t) \\[3pt] \boldsymbol{p}(t) \end{matrix} \right) = \sum_{i=1}^{\NN} \left[\lambda_i(t) \left( \begin{matrix} \boldsymbol{R}_i \\[3pt] \boldsymbol{P}_i \end{matrix} \right) \mathrm{e}^{i\omega_i t} + \lambda_i(t)^* \left( \begin{matrix} \boldsymbol{R}_i^* \\[3pt] \boldsymbol{P}_i^* \end{matrix} \right) \mathrm{e}^{-i\omega_i t}\right].
\end{equation}
Time evolution of the coefficients $\lambda_i(t)$ is determined by the time dependence of the vector $\boldsymbol{g}(t)$ and is given by the ordinary differential equation
\begin{equation}
\frac{\rmd \lambda_i(t)}{\rmd t}=-m g_i(t),
\end{equation}
where $g_i(t)$ are the coefficients in the following expansion of the vector $\boldsymbol{g}(t)$ into the mode amplitudes
\begin{equation}
\left( \begin{matrix}
0 \\[3pt] \boldsymbol{g}(t) \end{matrix} \right) =\sum_{i=1}^{\NN}\left[g_i(t)\left( \begin{matrix} \boldsymbol{R}_i \\[3pt] \boldsymbol{P}_i \end{matrix} \right) + g_i(t)^* \left( \begin{matrix} \boldsymbol{R}_i^* \\[3pt] \boldsymbol{P}_i^* \end{matrix} \right)\right].
\end{equation}

A full classical analysis of linear systems in three dimensions was given before \cite{IBBTS,TSIBB}. One can easily generalize this analysis to an arbitrary number of dimensions. 

\section{Quantum dynamics of wave packets}
In the position representation of a quantum system, the wave function $\Psi(\boldsymbol{r},t)$ obeys the following Schr\"odinger equation
\begin{equation} \label{schrodeq}
i\hbar \partial_t \Psi(\boldsymbol{r},t) = \left(-\frac{\hbar^2}{2m} \boldsymbol{\nabla}^2 + \frac{\hbar}{i} \boldsymbol{r}\!\cdot\!\hat{\Omega}\!\cdot\!\boldsymbol{\nabla} + \frac{m}{2}\boldsymbol{r}\!\cdot\!\hat{V}\!\cdot\!\boldsymbol{r}+m\boldsymbol{g}(t)\!\cdot\!\boldsymbol{r} \right) \Psi(\boldsymbol{r},t).
\end{equation}

\subsection{Dynamics of Gaussian wave packets}
In the first step of our analysis, let us consider the dynamics of a state of a quantum-mechanical system described by a Gaussian wave function of the form
\begin{equation} \label{funkcjafalowa}
\Psi(\boldsymbol{r},t) = N(t) \mathrm{e}^{i\phi(t)/\hbar} \exp\left( -\frac{m}{2\hbar} \left(\boldsymbol{r}-\boldsymbol{R}(t)\right)\!\cdot\!\hat{K}(t)\!\cdot\!\left(\boldsymbol{r}-\boldsymbol{R}(t)\right) + \frac{i}{\hbar}\boldsymbol{r}\!\cdot\!\boldsymbol{P}(t)\right),
\end{equation}
where the matrix $\hat{K}(t)$ is of course symmetric and its real part is positive. This wave function should also be normalized. Since the Schr\"odinger equation (\ref{schrodeq}) preserves normalization in time, it is enough to assume the normalization condition at the initial moment $t=0$
\begin{equation}
1 = \int_{\mathbb{R}^\NN} \rmd ^\NN\boldsymbol{r} |\Psi(\boldsymbol{r},0)|^{2}  = N(0)^2 \left[\frac{2\pi\hbar}{m \mathrm{Det}(\Re\hat{K}(0))}\right]^{\NN/2}.
\end{equation}
Parameters $\boldsymbol{R}$ and $\boldsymbol{P}$ have a direct interpretation as the position and momentum of the center of the wave function and at the same time they are the expectation values of the quantum operators of the position and momentum. 

One can show that the Schr\"odinger equation (\ref{schrodeq}) is equivalent to the following equations for the parameters of the wave function
\begin{subequations} \label{ewolucjaparametrow}
\begin{eqnarray}
\frac{\rmd \hat{K}(t)}{\rmd t} &=& -i\hat{K}(t)^2 + i\hat{V}-\left[\hat{\Omega},\hat{K}(t)\right], \label{shape}\\
\frac{\rmd \boldsymbol{R}(t)}{\rmd t} &=& \frac{\boldsymbol{P}(t)}{m}-\hat{\Omega}\!\cdot\!\boldsymbol{R}(t), \label{CMR}\\
\frac{\rmd \boldsymbol{P}(t)}{\rmd t} &=& -m\hat{V}\!\cdot\!\boldsymbol{R}(t)-\hat{\Omega}\!\cdot\!\boldsymbol{P}(t) -m\boldsymbol{g}(t), \label{CMP} \\
\frac{\rmd N(t)}{\rmd t} &=& \frac{N(t)}{2}\mathrm{Tr}(\Im \hat{K}(t)), \label{normalconst}\\
\frac{\rmd \phi(t)}{\rmd t} &=&-\frac{\hbar}{2}\mathrm{Tr}(\Re \hat{K}(t)) - \frac{\boldsymbol{P}(t)^2}{2m}+\frac{m}{2}\boldsymbol{R}(t)\!\cdot\!\hat{V}\!\cdot\!\boldsymbol{R}(t). \label{phaseevol}
\end{eqnarray}
\end{subequations}

Comparing the equations (\ref{CMR}) and (\ref{CMP}) with the classical equations of motion (\ref{eqmotion}), one can see that the dynamics of the center of the wave packet is the same as the dynamics of a classical particle. It is a well known realization of the Ehrenfest theorem which is, as we have said before, exactly satisfied for linear systems.

It is worth while to notice that the dynamics of the center of our wave packet is totally separated from the dynamics of the internal motion of the packet. This is a general property of quantum dynamics in external harmonic potentials \cite{IBB}. In addition, the evolution of the shape of our wave packet (described by a complex matrix $\hat{K}$) is not influenced by the external field $\boldsymbol{g}(t)$. The only place where the external field appears in our equations is the dynamics of the center of mass. Therefore, to solve the general problem of the motion of the Gaussian wave packet, one can solve at first a simpler problem of quantum dynamics without an external force and without motion of the center of the packet. Then one just needs to move this solution along a chosen classical trajectory. 

\subsection{Time evolution of the shape of the wave packet} 
The equation (\ref{shape}) describes the dynamics of the shape of the Gaussian state and it has the form of a well known matrix Riccati equation. Mathematical theory of these equations is well developed \cite{Reid,Lancaster}. In particular, the matrix Riccati equations find numerous applications in control theory \cite{Goodwin}. It is known that these nonlinear equations can be replaced by the linear ones. Following the standard procedure \cite{Reid}, we shall search for solutions of Eq. (19a) in the form
\begin{equation} \label{Kdef}
\hat{K}(t) = -\frac{i}{m} \hat{N}(t)\!\cdot\!\hat{D}^{-1}(t).
\end{equation}
This matrix equation is satisfied (as is shown in the Appendix \ref{appmatrix}) when the matrices $\hat{N}$ and $\hat{D}$ obey the following {\sl linear equations}
\begin{subequations} \label{linmatrixeq}
\begin{eqnarray}
\frac{\rmd \hat{N}}{\rmd t} &=& -m\hat{V}\!\cdot\!\hat{D}-\hat{\Omega}\!\cdot\!\hat{N}, \\
\frac{\rmd \hat{D}}{\rmd t} &=& \frac{1}{m}\hat{N}-\hat{\Omega}\!\cdot\!\hat{D}.
\end{eqnarray}
\end{subequations}
The linearization of the Riccati equation, in addition to being an effective mathematical tool, has also conceptual advantages. Namely, it leads to a direct relationship between classical and quantum theory. Comparing Eqs. (\ref{linmatrixeq}) with Eqs. (\ref{eqmotion}), one can see that the columns of the matrices $\hat{N}$ and $\hat{D}$ satisfy the same equations as the classical position and momentum vectors, respectively. Therefore, from the knowledge of the classical motion, one may determine the evolution of Gaussian wave function. It is a desirable manifestation of an exact connection between classical and quantum mechanics in the language of wave functions. 

\subsubsection{Example in one dimension} \label{przyklad1d}
Let us show now how this formalism works for the one dimensional standard harmonic oscillator \cite{Arnaud,Castro}. In this case our system is described by the Hamiltonian
\begin{equation}
{\cal H} = \frac{p^2}{2m} + \frac{m\omega^2}{2}x^2.
\end{equation}
In this case, the equations of the classical motion have a simple, scalar form
\begin{equation} 
\dot{x} = \frac{\partial {\cal H}}{\partial p} = \frac{p}{m}, \qquad \dot{p} = -\frac{\partial {\cal H}}{\partial x} = -m\omega^2 x.
\end{equation}
The solution of these equations satisfying the initial conditions $x(0)=x_0$, $p(0)=p_0$ is the following
\begin{subequations} \label{solclass}
\begin{eqnarray}
x(t) &=& x_0 \cos(\omega t) + \frac{p_0}{m\omega}\sin(\omega t), \\
p(t) &=& p_0 \cos(\omega t) - m\omega x_0 \sin(\omega t).
\end{eqnarray}
\end{subequations}
In the quantum mechanical case, the dynamics is described by the Schr\"odinger equation
\begin{equation} 
i\hbar \partial_t \Psi(x,t) = \left( -\frac{\hbar^2}{2m}\frac{\partial^2}{\partial x^2}+m\omega^2 x^2\right) \Psi(x,t).
\end{equation}
Let us assume that the wave function of the system has a Gaussian form
\begin{equation}
\Psi(x,t) = \left(\frac{\hbar \pi}{m\Re\alpha(t)}\right)^{\frac{1}{4}}\mathrm{e}^{i\phi(t)}\mathrm{e}^{-\frac{m}{2\hbar}\alpha(t)x^2},
\end{equation}
where the parameter $\alpha(t)$ describes the shape of our Gaussian packet. According to the equation (\ref{phaseevol}), the phase of our wave function is determined by the evolution the of $\alpha(t)$
\begin{equation}
\phi(t) = -\frac{1}{2}\int_0^t\Re \alpha(\tau) \rmd \tau.
\end{equation}

If we know that at the beginning the shape of our packet is given by the condition 
\begin{equation} \label{warunekal}
\alpha(0)=k\omega,
\end{equation}
we can predict what will be its value at any moment. The parameter $\alpha(t)$ obeys the following Riccati differential equation
\begin{equation}
\dot{\alpha}(t) = -i\alpha^2(t) + i\omega^2.
\end{equation}
Due to the Riccati procedure outlined before, we can linearize this equation by a substitution $\alpha=-in(t)/md(t)$ and then we know that the equations of the evolution of the parameters $d(t)$ and $n(t)$ are equivalent to the classical equations of motion (\ref{eqmotion})
\begin{equation} 
\dot{d}(t) = \frac{n(t)}{m}, \qquad \dot{n}(t) = -m\omega^2 d(t).
\end{equation}
Let us choose the initial conditions for $n(t)$ and $d(t)$ which are in accordance with (\ref{warunekal}) as follows
\begin{equation}
d(0) = -i\sqrt{\frac{\hbar}{m\omega}}, \qquad n(0)=k \sqrt{\hbar m \omega}.
\end{equation}
From the solution of the classical equations of motion (\ref{solclass}) one finds the time dependence of $d(t)$ and $n(t)$
\begin{subequations}
\begin{eqnarray}
d(t) &=& -i\sqrt{\frac{\hbar}{m\omega}}\left[ \cos(\omega t) - i k \sin(\omega t)\right], \\
n(t) &=& \sqrt{\hbar m \omega}\left[ k\cos(\omega t) -i\sin(\omega t) \right].
\end{eqnarray}
\end{subequations}
It means that the solution of the Schr\"odinger equation has a form
\begin{equation}
\Psi(x,t) = \left(\frac{\pi\hbar}{m\omega}\right)^{\frac{1}{4}}\left(\Re\frac{\cos(\omega t) - i k \sin(\omega t)}{k\cos(\omega t) -i\sin(\omega t)}\right)^{\frac{1}{4}}\mathrm{e}^{i\phi(t)}\exp\left(-\frac{m\omega}{2\hbar}\frac{k\cos(\omega t) -i\sin(\omega t)}{\cos(\omega t) - i k \sin(\omega t)}x^2\right).
\end{equation}
Such a wave function describes pulsating states known from the quantum mechanics courses. 

One should notice here that if  we put $k=1$ at the beginning of motion, then the time dependence will appear only in the phase of the wave function. It means that such a function is a stationary Gaussian wave packet of our system described by the Schr\"odinger equation (\ref{schrodeq}).
\section{Gaussian stationary quantum states}
As we have shown above in the one dimensional example using the Riccati method we can not only predict the evolution of the shape but we can also find a stationary state of our system whose shape is constant in time. We will show now how to obtain such a stationary Gaussian state in a general $\NN$ dimensional case. 

\subsection{Conditions of stationarity}
Since the evolution equations of the parameters $\boldsymbol{R}(t)$ and $\boldsymbol{P}(t)$ are completely separated from the rest, the Gaussian state can be stationary only when they are identically equal to zero. The condition that the shape is constant in time has the form
\begin{equation}
\frac{\rmd }{\rmd t}\hat{K}=0.
\end{equation}
This condition guarantees that the normalization factor $N(t)$ is constant in time, because it depends only on the shape of the wave packet. Therefore, as it is seen from the equation (\ref{normalconst}), the imaginary part of the matrix $\hat{K}$ is traceless. 

\subsection{Solving the algebraic matrix Riccati equation}
To find a stationary Gaussian state

\begin{equation} \label{stationarygauss}
\Psi_0(\boldsymbol{r}) \sim \exp\left[-\frac{m}{2\hbar}\boldsymbol{r}\cdot\hat{K}_0\cdot\boldsymbol{r}\right]
\end{equation}

we have to solve the following algebraic matrix Riccati equation
\begin{equation} \label{algriccati}
0 = -i\hat{K}_0^2 + i\hat{V}-\left[\hat{\Omega},\hat{K}_0\right],
\end{equation}
where the matrix $\hat{K}_0$ describes the shape of a stationary Gaussian state.

It is worth to notice that if we find two matrices $\hat{D}(t)$ and $\hat{N}(t)$ which satisfy the equations (\ref{linmatrixeq}) and have a following form
\begin{eqnarray}
\hat{D}(t) &=& \hat{D}_0 \!\cdot\! \hat{E}(t), \\
\hat{N}(t) &=& \hat{N}_0 \!\cdot\! \hat{E}(t).
\end{eqnarray}
then the matrix $\hat{K}_0$ defined by the equation analogous to (\ref{Kdef}) 
\begin{equation} \label{K0def}
\hat K_0 = -\frac{i}{m}\hat{N}_0\cdot\hat{D}_0^{-1}
\end{equation}
is a solution of the equation (\ref{algriccati}).

The problem of finding such matrices is not a hard task. After all, we know that the columns of these matrices obey exactly the same equations as the classical positions and momenta. Therefore, if we take the classical eigenmodes as the columns of these matrices, then the matrix $\hat E(t)$ simply will be a diagonal matrix with the elements $e^{i \omega_i t}$ and the matrices $\hat D_0$ and $\hat N_0$ can be build from the amplitudes $\boldsymbol{R}_i$ and $\boldsymbol{P}_i$ of the classical modes. 

\subsection{Proper choice of classical eigenmodes}
The construction of the matrix $\hat{K}_0$ outlined above is not yet completed. Still it is not clear which $\NN$ from $2\NN$ classical eigenmodes one should use in this construction. What are the requirements are necessary to make a good choice? The answer is: one should use such a set of classical modes that create a matrix $\hat{K}_0$ with a positive real part. It is a necessary requirement because we want this matrix to describe a real Gaussian, square integrable wave packet. Of course it can happen, that this construction is impossible, for example when one pair of characteristic frequencies is not real. In this case, our system does not have integrable quantum states but the construction of a stationary (unphysical) state is still possible. 

We should also notice here that even if the Hamiltonian is not bounded from below it can still be possible to construct integrable quantum stationary states of our system. 

\section{Other stationary states} \label{innestany}
\subsection{Evolution of wave packets with constant shapes}
It follows from the equations (\ref{ewolucjaparametrow}) that the motion of a Gaussian state with a constant shape $\hat{K}(t)=\hat{K}_0$, centered on the classical trajectory, is obviously possible. The wave function which describes such a situation has the form
\begin{equation} \label{gaussnalinie}
\Psi(\boldsymbol{r},t)=N\mathrm{e}^{i\phi(t)/\hbar}\mathrm{exp}\left[ -\frac{m}{2\hbar}(\boldsymbol{r}-\boldsymbol{R}(t))\!\cdot\!\hat{K}_0\!\cdot\!(\boldsymbol{r}-\boldsymbol{R}(t))+\frac{i\boldsymbol{P}(t)\!\cdot\!\boldsymbol{r}}{\hbar}\right].
\end{equation}
where the vectors $\boldsymbol{R}(t)$ and $\boldsymbol{P}(t)$ obey the classical equations of particle motion (\ref{eqmotion}). Therefore, let us choose as a solution the most general physical trajectory generated by all possible modes of the system (\ref{physsolution})
\begin{subequations}
\begin{eqnarray}
\boldsymbol{R}(t) &=& \sum_{i=1}^\NN\left(\lambda_i\boldsymbol{R}_i \mathrm{e}^{i\omega_i t}+\lambda_i^* \boldsymbol{R}_i^* \mathrm{e}^{-i\omega_i t}\right), \\
\boldsymbol{P}(t) &=& \sum_{i=1}^\NN\left(\lambda_i\boldsymbol{P}_i \mathrm{e}^{i\omega_i t}+\lambda_i^*\boldsymbol{P}_i^* \mathrm{e}^{-i\omega_i t}\right).
\end{eqnarray}
\end{subequations}
Set of coefficients $\lambda_i$ defines exactly one physical trajectory along which our gaussian solution with stationary shape moves. Wave function (\ref{gaussnalinie}) is just a multidimensional generalization of well known Glauber or coherent state. 

Without loss of generality one can assume that all modes used during construction of matrix $K_0$ are labeled as  $\boldsymbol{R}_i$ and $\boldsymbol{P}_i$. All other modes are complex conjuncted to them. 

Because $\boldsymbol{R}_i$ and $\boldsymbol{P}_i$ are vectors used during construction matrix $\hat{K}_0$ therefore from (\ref{K0def}) the following relations hold
\begin{subequations} \label{rel1}
\begin{eqnarray} 
\hat{K}_0\cdot \boldsymbol{R}_i&=&-\frac{i}{m}\boldsymbol{P}_i, \\
\hat{K}_0^*\cdot \boldsymbol{R}_i^*&=&\frac{i}{m}\boldsymbol{P}_i^*.
\end{eqnarray}
\end{subequations}

In addition from the equations (\ref{charpoly}) follows that for the amplitude vectors $\boldsymbol{R}_i$ and $\boldsymbol{P}_i$ of any mode of the system following relations hold
\begin{subequations} \label{rel2}
\begin{eqnarray} 
\boldsymbol{R}_i^*\cdot\boldsymbol{P}_j&=&\boldsymbol{R}_j\cdot\boldsymbol{P}_i^*,\qquad \mathrm{for}\qquad i\neq j \\
\boldsymbol{R}_i^*\cdot\boldsymbol{P}_i&=&-\boldsymbol{R}_i\cdot\boldsymbol{P}_i^*, \\
\boldsymbol{P}_i &=& m\left(\hat{\Omega}+i\omega_i\right)\cdot \boldsymbol{R}_i, \\
im\left(\omega_i+\omega_j\right)\boldsymbol{R}_i\cdot\boldsymbol{P}_j&=&\boldsymbol{P}_i\cdot\boldsymbol{P}_j-m^2\boldsymbol{R}_i\cdot\hat{V}\cdot\boldsymbol{R}_j,\\
im\left(\omega_i-\omega_j\right)\boldsymbol{R}_i\cdot\boldsymbol{P}_j^*&=&\boldsymbol{P}_i\cdot\boldsymbol{P}_j^*-m^2\boldsymbol{R}_i\cdot\hat{V}\cdot\boldsymbol{R}_j^*.
\end{eqnarray}
\end{subequations}

Using relations (\ref{rel1}) and (\ref{rel2}) one can easily show that the wave function (\ref{gaussnalinie}) can be represented in the following way
\begin{equation} \label{generujacafunkcja}
\Psi(\boldsymbol{r},t) = N\mathrm{e}^{i\phi_0}\mathrm{e}^{-i\Omega_0 t}\mathrm{exp}\left[-\sum_{i=1}^{\NN}\sum_{j=1}^{\NN} \frac{1}{2}\lambda_i^*\lambda_j^*\,\boldsymbol{R}_i^*\!\cdot\!\hat{A}\!\cdot\!\boldsymbol{R}_j^*e^{-i(\omega_i+\omega_j)t} +\sum_{i=1}^{\NN}\lambda_i^* \boldsymbol{r}\!\cdot\!\hat{A}\!\cdot\!\boldsymbol{R}_j^*e^{-i\omega_i t}\right]\!\Psi_0(\boldsymbol{r}),
\end{equation}
where
\begin{subequations}
\begin{eqnarray}
\hat{A} &=& \hat{K}_0+\hat{K}_0^*, \\
\Omega_0 &=& \frac{1}{4}\,\mathrm{Tr}(\hat{A}),
\end{eqnarray}
\end{subequations}
and $\phi_0$ is an arbitrary global phase factor.

From the definitions of modes vectors $\boldsymbol{R}_i$ and $\boldsymbol{R}_i^*$ it follows that any $\NN$ from $2\NN$ such vectors is a base in the $\NN$-dimensional space. Therefore we can represent any $\NN$ dimensional vector as a linear combination of chosen vectors. It means that there exist such coefficients $x_i$ that
\begin{equation}
\boldsymbol{r}=\sum_{i=1}^{\NN} x_i(\boldsymbol{r}) \boldsymbol{R}_i^*.
\end{equation}
To make next steps more clear let us introduce new quantities
\begin{subequations} \label{defpomoc}
\begin{eqnarray}
\gamma_{ij}&=& \frac{m}{\hbar}\boldsymbol{R}_i^*\cdot\hat{A}\cdot\boldsymbol{R}_j^*, \label{gammaij} \\
z_i&=& \lambda_i^* e^{-i\omega_i t}.
\end{eqnarray}
\end{subequations}

From the definition (\ref{gammaij}) one has the symmetry property $\gamma_{ij}=\gamma_{ji}$.

Using (\ref{defpomoc}) one can represent the wave function of the coherent state (\ref{generujacafunkcja}) as follows
\begin{equation} \label{funkcjaok}
\Psi(\boldsymbol{r},t) = N\mathrm{e}^{i\phi_0}\mathrm{e}^{-i\Omega_0 t}\mathrm{exp}\left[-\frac{1}{2}\sum_{i=1}^{\NN}\sum_{j=1}^{\NN} \gamma_{ij}z_iz_j +\sum_{i=1}^{\NN}\sum_{j=1}^{\NN} \gamma_{ij}x_iz_j\right]\!\Psi_0(\boldsymbol{r}).
\end{equation}

This form of the wave function allows us to find all stationary states of the system.

\subsection{Expansion into stationary states} \label{rozdzialrozwiniecie}
To get an expansion of the wave function (\ref{funkcjaok}) into stationary states (eigenstates of the Hamiltonian of the system) we should use the concept of, so called multidimensional Hermite polynomials. Properties of these polynomials are well studied \cite{Herm1,Herm2} and a general information about them can be found in the Appendix \ref{appherm}. The generating function of $\NN$ dimensional Hermite polynomials $\mathrm{H}_{n_1,\ldots,n_\NN}^\gamma(x_1,\ldots,x_\NN)$ has a form
\begin{equation}
\mathrm{exp}\left[\sum_{i=1}^{\NN}\sum_{j=1}^{\NN} \gamma_{ij}(x_i-\frac{z_i}{2})z_j \right]=\sum_{(n_1,\ldots,n_\NN)=0}^\infty\frac{z_1^{n_1}}{n_1!}\cdots\frac{z_\NN^{n_\NN}}{n_\NN!}\mathrm{H}_{n_1,\ldots,n_\NN}^\gamma(x_1,\ldots,x_\NN)
\end{equation}
One can use this expansion in (\ref{funkcjaok}) and then our wave function becomes
\begin{equation} \label{porozw}
\Psi(\boldsymbol{r},t) = N\mathrm{e}^{i(\Omega_0t+\phi_0)}\!\!\!\!\!\!\sum_{(n_1,\ldots,n_\NN)=0}^\infty\left[\prod_{i=1}^{\NN} \frac{(\lambda_i^*)^{n_i}}{n_i!}e^{-in_i\omega_it}\right]\mathrm{H}_{n_1,\ldots,n_\NN}^{\gamma}\!\left(x_1,\ldots,x_\NN\right) \,\Psi_0(\boldsymbol{r}).
\end{equation}

Therefore our wave function is a superposition of the stationary states --- the states whose evolution in time appears only in the evolution of the phase. For each set $(n_1,\ldots,n_\NN)$ we have defined exactly one (up to the normalization factor) stationary state --- the state which arise form the ground state by the excitations in the modes ($i$-th mode is excited $n_i$ times)
\begin{equation} 
\Psi_{n_1,\ldots,n_\NN}(\boldsymbol{r}) \sim \mathrm{H}_{n_1,\ldots,n_\NN}^\gamma\left(x_1,\ldots,x_\NN\right) \Psi_0(\boldsymbol{r}).\label{stanstac}
\end{equation}

It is worth to notice that the only place where the scale factors $\lambda_i$ appear in (\ref{porozw}) is the coefficient standing before the Hermite polynomial. Therefore, the wave function (\ref{stanstac}) does not depend on it. It means that set of the stationary states is independent on trajectory used during construction.

It is obvious  from the Schr\"odinger equation that the eigenvalue of the Hamiltionian in the state described by wave function (\ref{stanstac}) is equal \mbox{$\hbar\left( \sum_{i=1}^{\NN}n_i\omega_i + \Omega_0\right)$}.

All states with all possible sets of $(n_1,\ldots,n_\NN)$ form a base in the wave-function space. Because of the properties (\ref{Horth}) and (\ref{Hrozklad}) of the multidimensional Hermite polynomials the completeness of this set is guaranteed and any quantum state of the system can be represented as a superposition of these states
\begin{equation}
\psi(\boldsymbol{r},t) = \!\!\!\!\sum_{(n_1,\ldots,n_\NN)=0}^\infty\!\!\!\!C_{n_1,\ldots,n_\NN}(t) \Psi_{n_1,\ldots,n_\NN}(\boldsymbol{r}).
\end{equation}

At this moment it is worth while to notice that the construction outlined above uses effectively only this modes which were not used during the construction of the matrix $\hat{K}_0$. One can easily see that amplitudes $\boldsymbol{R}_i$ do not play any role in the construction. It means that we have some kind of division of roles. Half of the modes is responsible for the creation of a positive defined matrix $\hat{K}_0$ and the ground state of the system. While the other half gives the stationary state of the system.

This observation has a very nice physical interpretation. When the Hamiltonian of the system is bounded from below, one can show that the proper classical modes (which guarantee that the matrix $\hat{K}_0$ is positive defined) have positive frequencies. Therefore, when we create excited states, we have to use the modes with negative frequencies and their energy (eigenvalue of the Hamiltonian) is growing with $n_i$. On the other hand, when the Hamiltonian is not bounded from below, then some of the modes contained in the matrix $\hat{K}_0$ have negative frequencies. Therefore, when exciting one of these modes, one uses a mode with positive frequency and the energy of the generated states diminishes --- it is not bounded from below. 

\section{Conclusions}
In this paper we have shown how the classical-quantum correspondence for the linear systems is represented in the language of wave functions. The dynamics of Gaussian states is completely determined by the dynamics of a classical particle in a very unexpected way --- the shape of Gaussian state evolves like a quotient of the classical momentum and position. This observation enables us to construct a stationary Gaussian state of the system. In addition the classical dynamics of the center of the Gaussian state, leads to a complete set of stationary states. 

\section{Acknowledgments}
The author would like to thank prof. Iwo Bia\l ynicki-Birula for his fruitful discussions and important questions which have changed author's point of view on this topic. 

This research was supported by a grant from the Polish Committee for Scientific Research in the years 2004-06.

\appendix
\section{Two mathematical theorems} \label{appmatrix}
\subsection{About the matrix Riccati equation}
The matrix Riccati equation in $n$ dimensions has the form
\begin{equation} \label{riccatiequation}
\frac{\rmd \hat{K}(t)}{\rmd t} = \alpha \hat{K}^2(t) + \hat{K}(t)\!\cdot\!\hat{W} + \hat{W}^{\mathrm{T}}\!\cdot\!\hat{K}(t) + \hat{U}.
\end{equation}
where $\hat{W}$ and $\hat{U}$ are $n \times n$ matrices and  $\alpha$ is a constant parameter. The Riccati theorem \cite{Reid} says that $\hat{K}(t)$ is a solution of the equation (\ref{riccatiequation}) if and only if there exist matrices $\hat{A}(t)$ and $\hat{B}(t)$ which satisfy the following equations
\begin{subequations} \label{ukladriccati}
\begin{eqnarray} 
\frac{\rmd \hat{A}(t)}{\rmd t} &=& \hat{U}\!\cdot\!\hat{B}(t) + \hat{W}^{\mathrm{T}}\!\cdot\!\hat{A}(t), \label{u1}\\
\frac{\rmd \hat{B}(t)}{\rmd t} &=& -\alpha\hat{A}(t) - \hat{W}\!\cdot\!\hat{B}(t). \label{u2}
\end{eqnarray}
\end{subequations}
and then the solution has the form
\begin{equation} \label{defrozwiazania}
\hat{K}(t) = \hat{A}(t)\!\cdot\!\hat{B}^{-1}(t).
\end{equation}
It is obvious, that if there exist matrices $\hat{A}$ and $\hat{B}$ which satisfy the equations (\ref{ukladriccati}) then the matrix $\hat{K}(t)$ defined by an equation (\ref{defrozwiazania}) is a solution of the Riccati equation (\ref{riccatiequation}). On the other hand, if $\hat{K}(t)$ obeys the equation (\ref{riccatiequation}) then one can define the matrices $\hat{A}(t)$ and $\hat{B}(t)$
\begin{subequations} 
\begin{eqnarray}
\hat{B}(t) &=& \mathrm{T}\exp \left[-\int_0^t \left(\alpha \hat{K}(\tau)+\hat{W}\right) \rmd \tau \right], \\
\hat{A}(t) &=& \hat{K}(t) \!\cdot\! \hat{B}(t). \label{rownKodwr}
\end{eqnarray}
\end{subequations}
One can then show that such matrices obey the equations (\ref{ukladriccati}).
\subsection{About the eigenvalues of some matrix}
Let us define $2n \times 2n$ matrix $\hat{\cal M}$ as follows
\begin{equation} \label{macierzM}
\hat{\cal M} = \begin{pmatrix} \hat{W} & \hat{A} \\ \hat{B} & \hat{W}^{\mathrm{T}}\end{pmatrix}
\end{equation}
where $\hat{W}$, $\hat{A}$, $\hat{B}$ are real $n\times n$ matrices and $\hat{A}$ and $\hat{B}$ are symmetric. Then if $\lambda$ is an eigenvalue of the matrix (\ref{macierzM}) then $-\lambda$ is an  eigenvalue of this matrix too. It means that the characteristic polynomial is of the order $n$ in $\lambda^2$. This theorem follows from the observation that the matrices $\hat{\cal M}$ and $-\hat{\cal M}$ have the same set of eigenvalues because one has following reflexive property
\begin{equation}
-\hat{\cal M}^{\mathrm{T}} = \hat{\cal S}\!\cdot\!\hat{\cal M}\!\cdot\!\hat{\cal S}^{-1}, \qquad 
\hat{\cal S} = \begin{pmatrix} 0 & \hat{I} \\ -\hat{I} & 0 \end{pmatrix}.
\end{equation}
\section{Multidimensional Hermite polynomials} \label{appherm}

Multidimensional Hermite polynomials used in section \ref{rozdzialrozwiniecie} were introduced in \cite{Herm1} and deep analysis of their properties was studied in \cite{Herm2}. Nowadays these polynomials are often used in quantum optics contects \cite{Opt1,Opt2} in description of mixed and entangled states of electromagnetic filed. 

Multidimensional Hermite polynomials in $\NN$ dimensions are functions of $\NN$ variables ${x_1,\ldots,x_\NN}$. Let us assume that $\hat{\gamma}$ is symmetric and positive $\NN\times\NN$ matrix. Then we define $\NN$ dimensional Hermite polynomial of the order $(n_1, n_2, \ldots, n_\NN)$ as follows
\begin{equation}
\mathrm{H}_{n_1,\ldots,n_\NN}^{\gamma}\left(x_1,\ldots,x_\NN\right) = (-1)^{\sum_{i=1}^\NN n_i}\, e^{\frac{1}{2}\boldsymbol{x}\cdot\hat{\gamma}\cdot\boldsymbol{x}}\frac{\partial^{n_1}}{\partial x_1\,^{n_1}}\cdots\frac{\partial^{n_\NN}}{\partial x_\NN\,^{n_\NN}}\,e^{-\frac{1}{2}\boldsymbol{x}\cdot\hat{\gamma}\cdot\boldsymbol{x}}.
\end{equation}
In one dimensional case it is a set of well known Hermite polynomials defined as
\begin{equation}
\mathrm{H}_n(x) = (-1)^n e^{x^2}\frac{\rmd ^n}{\rmd x^n}e^{-x^2}.
\end{equation}

In analogy to standard Hermite polynomials multidimensional Hermite polynomials can be also constructed with using a generating function which has a following form
\begin{equation}
\mathrm{exp}\left[\sum_{i=1}^{\NN}\sum_{j=1}^{\NN} \gamma_{ij}\left(x_i-\frac{z_i}{2}\right)z_j\right]=\!\!\!\!\!\sum_{(n_1,\ldots,n_\NN)=0}^{\infty}\! \frac{z_1^{n_1}}{n_1!}\cdots\frac{z_\NN^{n_\NN}}{n_\NN!}\,\mathrm{H}_{n_1,\ldots,n_\NN}^{\gamma}\left(x_1,\ldots,x_\NN\right).
\end{equation}

It is just a generalization of well known generating function of Hermite polynomials
\begin{equation}
\mathrm{exp}\left[2z\left(x-\frac{z}{2}\right)\right] = \sum_{n=0}^{\infty}\frac{z^n}{n!}\mathrm{H}_n(x).
\end{equation}

Properties of the multidimensional Hermite polynomials are well studied \cite{Herm1,Herm2}. It was proved that they are orthogonal in the following sense
\begin{equation} \label{Horth}
\int_{\mathbb{R}^\NN} \rmd ^\NN\boldsymbol{x}\, e^{-\frac{1}{2}\boldsymbol{x}\cdot\hat{\gamma}\cdot\boldsymbol{x}}\,\mathrm{H}_{n_1,\ldots,n_\NN}^\gamma\left(\boldsymbol{x}\right)\mathrm{H}_{m_1,\ldots,m_\NN}^\gamma\left(\boldsymbol{x}\right) = \pi^{\NN/2}\prod_{i=1}^{\NN}2^{n_i} n_i!\delta_{n_im_i}.
\end{equation}
and any function $f(x_1,\ldots,x_\NN)$ of the class $L_w^2$, that is to say function for which the integral
\begin{equation}
\int_{\mathbb{R}^\NN} \rmd ^\NN\boldsymbol{x}\, e^{-\frac{1}{2}\boldsymbol{x}\cdot\hat{\gamma}\cdot\boldsymbol{x}} \left|f(x_1,\ldots,x_\NN)\right|
\end{equation}
is convergent, can be expanded in series of $\NN$ dimensional Hermite polynomials \cite{Herm3}
\begin{equation} \label{Hrozklad}
f(x_1,\ldots,x_\NN) = \sum_{(n_1,\ldots,n_\NN)=0}^\infty f_{n_1,\ldots,n_\NN} \mathrm{H}_{n_1,\ldots,n_\NN}^\gamma\left(x_1,\ldots,x_\NN\right)
\end{equation}
where 
\begin{equation}
f_{n_1,\ldots,n_\NN}=\left(\prod_{i=1}^{\NN}n_i!\right)^{-1}\int_{\mathbb{R}^\NN} \rmd ^\NN\boldsymbol{x}\, e^{-\frac{1}{2}\boldsymbol{x}\cdot\hat{\gamma}\cdot\boldsymbol{x}} f(x_1,\ldots,x_\NN)\,\mathrm{H}_{n_1,\ldots,n_\NN}^\gamma\left(x_1,\ldots,x_\NN\right).
\end{equation}
Many other properties of the multidimensional Hermite polynomials may be found (cf. for example \cite{Herm2}).


\begin{thebibliography}{99}
\bibitem{Ehrenfest} P. Ehrenfest, {\em Bemerkung \"uber die ange\"aherte G\"ultigkeit der klassischen Mechanik innerhalb der Quantenmechanik}, Z. Phys.~\textbf{45}, 455-457 (1927)
\bibitem{IBBTS} I. Bia\l ynicki-Birula and T. Sowi\' nski, {\em Gravity-induced resonances in a~rotating trap}, Phys. Rev. A~\textbf{71}, 043610 (2005).
\bibitem{Andrews} M. Andrews, {\em Invariant operators for quadratic Hamiltonians}, Am. J. Phys.~\textbf{67} (4), 336-343 (1999)
\bibitem{TSIBB} T. Sowi\' nski and I. Bia\l ynicki-Birula, {\em Harmonic oscillator in a rotating trap: Complete solution in 3D}, quant-ph/\textbf{0409070}, (2005).
\bibitem{Reid} W. T. Reid {\em Riccati Differential Equations}, Academic Press, New York, (1972).
\bibitem{Goodwin} G. C. Goodwin, S. F. Graebe, and M. E. Salgado, {\em Control System Design}, Prentice-Hall, Englewood Cliffs, Mass. (2001), Appendix D.
\bibitem{IBB} I. Bia\l ynicki-Birula and Z. Bia\l ynicka-Birula, {\em Center-of-mass motion in the many-body theory of Bose-Einstein condensates}, Phys. Rev. A~\textbf{65}, 063606 (2002).
\bibitem{Castro} A. S. de Castro and N. C. Cruz, {\em A pulsating Gaussian wave packet}, Eur. J. Phys.~\textbf{20}, L19-L20 (1999).
\bibitem{Arnaud} J. Arnaud, {\em Pulsating Gaussian wavepackets and complex trajectories}, Eur. J. Phys.~\textbf{21}, L15-L16 (2000).
\bibitem{Lancaster} P. Lancaster, {\em The role of the hamiltionian in the solution of algebraic Riccati equations}, Progr. Systems Control Theory, Integral Equations and Operator Theory, Vol. \textbf{25}, Birkh\"auser (1999), pages 157-172.
\bibitem{Herm1} P. Appell and J. Kamp´e de F´eriet, {\em Fonctions Hyperg\'eom\'etriques et Hypersph\'eriques}, Paris: Gauthier-Villars (1926)
\bibitem{Herm2} Erd\'elyi, {\em Bateman Manuscript Project: Higher Transcendental Functions}, New York: McGraw-Hill (1953)
\bibitem{Herm3} R. Caccippoli, {\em Giorn. Ist. Ital. Attuari} {\bf 3} 364-275 (1932)
\bibitem{Opt1} V. V. Dodonov, O. V. Man'ko, V. I. Man'ko {\em Multidimensional Hermite polynomials and photon distribution for polymode mixed light}, Phys. Rev. A~\textbf{50}, 1  (1994)
\bibitem{Opt2} P. Kok and S. L. Braunstein {\em Multi-dimensional Hermite polynomials in quantum
optics}, quant-ph/\textbf{0011114} (2000)
\end{thebibliography}
\end{document}